\documentclass[8.5pt,twoside,twocolumn]{article}
\usepackage[super,sort&compress,comma]{natbib}
\usepackage{mhchem}
\usepackage{times,mathptm}
\usepackage{sectsty}
\usepackage{balance}

\usepackage{graphicx} 
\usepackage{lastpage}
\usepackage[format=plain,justification=raggedright,singlelinecheck=false,font=small,labelfont=bf,labelsep=space]{caption}
\usepackage{fancyhdr}
\begin{document}

\thispagestyle{plain}
\fancypagestyle{plain}{
\fancyhead[L]{\includegraphics[height=8pt]{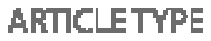}}
\fancyhead[C]{\hspace{-1cm}\includegraphics[height=20pt]{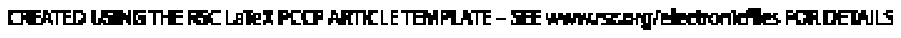}}
\fancyhead[R]{\includegraphics[height=10pt]{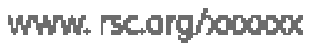}\vspace{-0.2cm}}
\renewcommand{\headrulewidth}{1pt}}
\renewcommand{\thefootnote}{\fnsymbol{footnote}}
\renewcommand\footnoterule{\vspace*{1pt}%
\hrule width 3.4in height 0.4pt \vspace*{5pt}}
\setcounter{secnumdepth}{5}

\makeatletter
\def\subsubsection{\@startsection{subsubsection}{3}{10pt}{-1.25ex plus -1ex minus -.1ex}{0ex plus 0ex}{\normalsize\bf}}
\def\paragraph{\@startsection{paragraph}{4}{10pt}{-1.25ex plus -1ex minus -.1ex}{0ex plus 0ex}{\normalsize\textit}}
\renewcommand\@biblabel[1]{#1}
\renewcommand\@makefntext[1]%
{\noindent\makebox[0pt][r]{\@thefnmark\,}#1}
\makeatother
\renewcommand{\figurename}{\small{Fig.}~}
\sectionfont{\large}
\subsectionfont{\normalsize}

\fancyfoot{}
\fancyfoot[LO,RE]{\vspace{-7pt}\includegraphics[height=9pt]{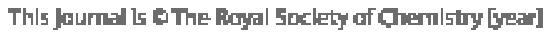}}
\fancyfoot[CO]{\vspace{-7.2pt}\hspace{12.2cm}\includegraphics{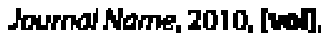}}
\fancyfoot[CE]{\vspace{-7.5pt}\hspace{-13.5cm}\includegraphics{headers/RF.ps}}
\fancyfoot[RO]{\footnotesize{\sffamily{1--\pageref{LastPage} ~\textbar  \hspace{2pt}\thepage}}}
\fancyfoot[LE]{\footnotesize{\sffamily{\thepage~\textbar\hspace{3.45cm} 1--\pageref{LastPage}}}}
\fancyhead{}
\renewcommand{\headrulewidth}{1pt}
\renewcommand{\footrulewidth}{1pt}
\setlength{\arrayrulewidth}{1pt}
\setlength{\columnsep}{6.5mm}
\setlength\bibsep{1pt}

\twocolumn[
\begin{@twocolumnfalse}
\noindent\LARGE{\textbf{Microphase separation in linear multiblock copolymers under poor solvent conditions}}
\vspace{0.6cm}

\noindent\large{\textbf{Panagiotis E.
Theodorakis,$^{\ast}$\textit{$^{a}$} and Nikolaos G.
Fytas\textit{$^{b}$}}}\vspace{0.5cm}

\noindent\textit{\small{\textbf{Received Xth XXXXXXXXXX 20XX, Accepted Xth XXXXXXXXX 20XX\newline
First published on the web Xth XXXXXXXXXX 200X}}}

\noindent \textbf{\small{DOI: 10.1039/b000000x}}
\vspace{0.6cm}

\noindent \normalsize{Molecular dynamics simulations are used to
study the phase behavior of linear multiblock copolymers with two
types of monomers, A and B, where the length of the polymer blocks
$N_{A}$ and $N_{B}$ ($N_{A}=N_{B}=N$), the number of the blocks
$n_{A}$ and $n_{B}$ ($n_{A}=n_{B}=n$), and the solvent quality
varies. The fraction $f$ of A-type monomers is kept constant and
equal to $0.5$. Whereas at high enough temperatures these
macromolecules form coil structures, where each block A or B forms
rather individual clusters, at low enough temperatures A and B
monomers from different blocks can join together forming clusters
of A or B monomers.  The dependence of the formation of these
clusters on the varied parameters is discussed in detail,
providing a full understanding of the phase behavior of linear
multiblock copolymers, at least for this symmetrical case.}
\vspace{0.5cm}
 \end{@twocolumnfalse}
  ]

\section{Introduction}
\label{sec:1}

\footnotetext{\textit{$^{a}$~Institut f\"ur Physik, Johannes
Gutenberg Universit\"at, Mainz, Germany. E-mail:
theodora@uni-mainz.de}} \footnotetext{\textit{$^{b}$~Department of
Materials Science, Patras 26504, Patras, Greece. }}

The phase separation behavior of block copolymers has been the
subject of several theoretical and experimental
studies.~\cite{1,2,3,4,5,6,7,8,9,10,11,12,13,14,15,16,17,18,19,20,21,22,23,24,25,26,27,28,29,30,31,32,33,34,35,36}
In the case of infinitely dilute solutions, it is sufficient to
evaluate the behavior of isolated chains. Very interesting is the
case of multiblock copolymers with blocks composed of either A- or
B-type monomers for which very successful methods of their
synthesis exist.~\cite{31,32} Apart from the experimental
characterization, numerical investigations of model chains are the
most direct approach to understanding the behavior of these
systems.~\cite{33,34,35,36} Interestingly, for the case of
multiblock copolymers there is also close relation to the various
toy-models (i.e., the HP model~\cite{37,38,39,40,41,42}), which try
to mimic the behavior of biomacromolecules on the way to
understanding complicated biological processes, i.e., protein
folding, helical structures~\cite{43}, etc.

Multiblock copolymers of two chemically different type of blocks
(A and B) are expected to adopt at high temperatures coil
structures, where chain conformations are essentially governed by
repulsive interactions between the different blocks. The result is
an expansion of the chain dimensions, not only with respect to the
unperturbed state, but also with respect to a homopolymer of the
same length under the same thermodynamic conditions.~\cite{44}
Also, one expects that the spherical symmetry of these
macromolecules should break, so that the chain forms a slightly
elongated object, but such phenomena will not be discussed in the
present manuscript. Here, we focus on the microphase separation
between the blocks of different type in a multiblock copolymer
chain with two types of blocks (A and B) and the dependence on its
structural parameters in terms of their cluster analysis. In this
direction, the most interesting properties of such macromolecules
are obtained when the solvent quality varies. Under poor solvent
conditions (temperatures lower than the Theta temperature of the
chain, which in our case will be the same for two different types
of blocks due to the symmetry of our model) the chain collapses
forming globular structures, where microphase separation between
different blocks takes place. The different blocks A and B come
together and form clusters of monomers of type A or B, which are
microphase separated. In this study, we analyze these
phenomena by studying the formation of these clusters and leave
the discussion of their size, in conjunction with the size of the
individual blocks, for a later communication. To this end, we have
performed large-scale molecular dynamics (MD) simulations of an
off-lattice model of a linear multiblock copolymer varying the
solvent quality to provide a first approach to the understanding
of phase separation of such macromolecules.

The outline of this paper is as follows. Sec.~\ref{sec:2} describes
our model and sketches the analysis needed to characterize the
size and overall shape (and other properties) of the clusters considering
average properties. Then, Sec.~\ref{sec:3} presents our numerical
results and Sec.~\ref{sec:4} summarizes our conclusions.

\section{Model and simulation methods}
\label{sec:2}

\begin{figure}[h]\centering
\includegraphics[height=0.60cm]{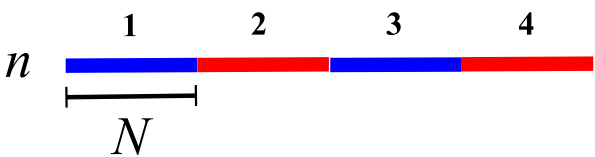}
\vspace{0.3in} \caption{Definition of structural parameters
describing our multiblock copolymer chains. $n$ is the number of
different blocks A and B (in this case $n=4$) and $N$ is the
length of each block. All the blocks, irrespective of whether they
are of type A or B, have the same length $N$. Then, the total
length of the chain is $nN$.} \label{fig:1}
\end{figure}

In this study we consider the most symmetric case of multiblock
copolymers, where the fraction of monomers of type A is $f=0.5$.
Then the length of blocks A, $N_{A}$, and blocks B, $N_{B}$, is
equal as well, and the number of blocks $n_{A}$ with monomers of
type A and $n_{B}$ with monomers of type B is the same. In this
study we denote the total number of blocks as $n$
($n=n_{A}+n_{B}$) and the length of the blocks with the capital
letter $N$ ($N=N_{A}=N_{B}$). Figure~\ref{fig:1} shows
schematically the definition of the above parameters. Then, the
total length of the multiblock chain is $nN$. Clearly, this choice
facilitates the study of this problem, being the most symmetric
case. Our chains are modelled by the standard bead-spring
model,~\cite{45,46,47} where all beads interact with a truncated
and shifted Lennard-Jones (LJ) potential
\begin{eqnarray}
\label{eq:1}
U_{LJ} (r) = \left\{ \begin{array}{l@{\; ,\;}l} 4
\epsilon_{LJ}[(\sigma _{LJ}/r)^{12}-(\sigma _{LJ}/r)^6 ] +C & r
\leq r_c\\
0 & r>r_c,
\end{array} \right.
\end{eqnarray}
where $r_c=2.5 \sigma _{LJ}$ is the cut-off of the potential, and
the constant $C$ is defined such that $U_{LJ}(r=r_c)$ is
continuous at this cut-off. Henceforth, units are chosen such that
$\epsilon_{LJ}=1$, $\sigma _{LJ}=1$, $k_B=1$, and $m=1$ (mass of
the beads) for simplicity. When we consider two types (A,B) of
blocks, we still use $\sigma_{LJ}^{AA} = \sigma_{LJ}^{AB} = \sigma
_{LJ}^{BB} = 1$ and $\epsilon_{LJ}^{AA}=\epsilon_{LJ}^{BB}=1$ but
$\epsilon _{LJ}^{AB}=1/2$ to create an unmixing tendency ($\Delta
\epsilon=\epsilon_{AB}-1/2(\epsilon_{AA}+\epsilon_{BB})$). We know
that in the case of a binary system with monomers at density
$\rho=1$ (e.g., a LJ mixture which is a standard system for the
study of phase separation), microphase separation occurs below a
critical temperature $T_c$ close to $T=1.5$.~\cite{48} For the
multiblock copolymers the average densities are much smaller, but
since the critical temperature scales proportional to the chain
length ($T_c \propto \chi_{crit}^{-1}$), we do expect to be able
to detect microphase separation with our model. For this we also
exploit previous experience with this model where microphase
separation was studied in the case of bottlebrush polymers with
two types of side chains.~\cite{51}

\vspace{0.3in}

The connectivity of the beads along the chain is maintained by the
finitely extensible nonlinear elastic (FENE) potential
\begin{equation}
\label{eq:2}
U_{FENE}(r) = - \frac 1 2 kr_0^2\ln [1-(r/r_0)^2]\;,0 <r \leq r_0,
\end{equation}
where the standard choice of parameters ($r_0=1.5$ and $k=30$) was
adopted, and $U_{FENE}(r>r_0)= \infty$.

\begin{figure}[h]\centering
\includegraphics[height=3cm]{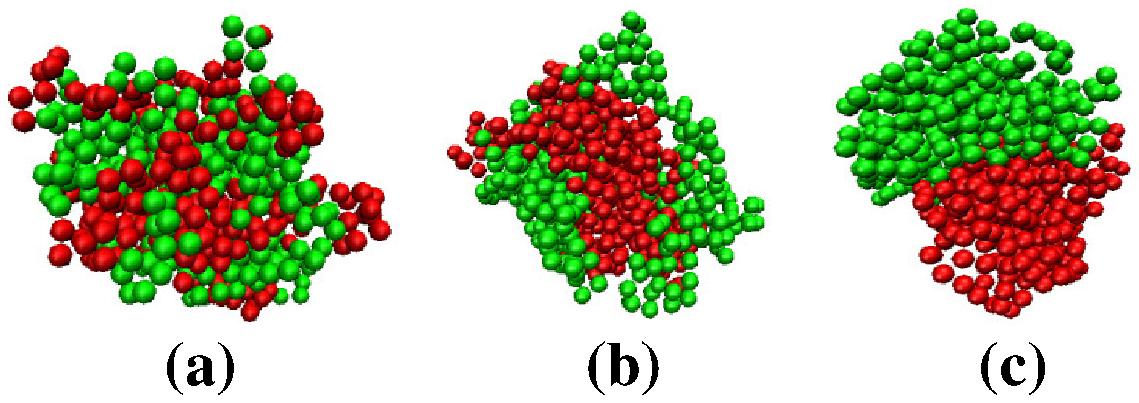}
\caption{Snapshots of three different multiblock copolymers of the
same total length $nN=600$ at a low temperature $T=1.5$ ((a):
$N=6$, (b): $N=15$, and (c): $N=60$). In case (c) we have the
formation of two clusters of different blocks which are always
phase separated.} \label{fig:2}
\end{figure}
For the model defined by Eqs.~(\ref{eq:1}) and (\ref{eq:2}) the
Theta temperature is known only rather roughly,~\cite{49} namely
$\Theta \approx 3.0$. Being interested in $T \leq \Theta$, we have
attempted to study the temperature range $1.5 \leq T \leq 3.0$.
However, as it is well-known, equilibration of collapsed chains is
a rather difficult task and our simulation procedure will be
discussed below. The temperature is controlled by the Langevin
thermostat, as done in previous work.~\cite{45,46,47,49} The
equation of motion for the coordinates $\{\vec{r}_i(t)\}$ of the
beads
\begin{equation}
\label{eq:3}
m \frac {d^2\vec{r}_i}{dt^2} = - \nabla U_i-m\gamma \frac
{d\vec{r}_i}{dt} + \vec{\Gamma}_i(t)
\end{equation}
is numerically integrated using the GROMACS package.~\cite{50} In
Eq.~(\ref{eq:3}) $t$ denotes the time, $U_i$ is the total
potential the $i$-th bead experiences, $\gamma$ is the friction
coefficient, and $\vec{\Gamma}_i(t)$ the random force. $\gamma$
and $\vec{\Gamma}_i(t)$ are related by the fluctuation-dissipation
relation as follows
\begin{equation}\label{eq:4}
\langle \vec{\Gamma}_i(t)\cdot \vec{\Gamma}_j(t')\rangle = 6 k_BT
\gamma \delta _{ij}\delta (t-t')\;.
\end{equation}
As in previous work,~\cite{45,46,47,49,51,52,53} the friction
coefficient was chosen $\gamma = 0.5$. For the integration of
Eq.~(\ref{eq:3}) the leap frog algorithm~\cite{54} was used with a
time step of $\Delta t = 0.006 \tau$, where the natural time unit
is defined as $\tau = (m \sigma ^2_{LJ}/\epsilon_{LJ})^{1/2}=1$.

As we mentioned earlier, equilibration of collapsed chains via MD
methods is difficult, and thus we briefly describe here our
equilibration procedure. First the system was equilibrated at
$T=3.0$ for a time range of $30\times 10^6 \tau$. To gather
statistics, a sufficient number of statistically independent
configurations (typically greater than $500$) at this temperature
was used for initial configurations of slow cooling runs, where
the temperature $T$ was lowered in steps of $0.1$, and running the
system at each $T$ for a time which exceeds the relaxation time of
our chains. The final configuration of each (higher) $T$ was used
as starting configuration for the next (lower) $T$. In this way,
we are able to generate statistically independent configurations
of clusters. One can see already from snapshot pictures of
Fig.~\ref{fig:2} the different structures that occur at a low
temperature $T=1.5$ for different multiblock chains, where $nN$ is
constant, but $N$ varies. Indeed, rather dense clusters
containing several blocks occur, and the conformation depends on
the parameters $N$, $n$, and $T$ (remember that $\chi N \propto
\Delta \epsilon / T$). We also note that as the length of the
chains increases, equilibration of the chains becomes extremely
difficult preventing the study of very long chains.
With our choice of
parameters, we expect to observe the most interesting effects in
the range of simulated chain lengths used in this study. In
particular, we study in detail the chain length $nN=600$ and then
we discuss the dependence on $n$ and $N$ separately, with $N$ of
course being the main parameter controlling the incompatibility
between A and B, for the chosen set of (potential) parameters.

When the chains collapse at the lower temperatures to form
microphase separated cluster structures, there is no reason a
priori for the system to decide which blocks will belong to a
specific cluster. We actually observe fluctuations where a block
of type A or B that was part of a cluster of similar monomers
escapes from the cluster to become part of another neighboring
cluster with blocks of the same type of monomers. In fact, many
such fluctuations would amount to influencing the characteristic
properties (number $N_{cl}$ of blocks a cluster contains, and its
linear dimensions) of each cluster.

For the analysis of the clusters one needs to identify for each
configuration that is analyzed, which blocks belong to which
cluster. We have used the standard Stillinger~\cite{55}
neighborhood criterion for monomers: if two monomers are less than
a distance $r_n$ apart, they belong to the same cluster. We
followed the standard choice $r_n=1.5 \sigma_{LJ}$ and checked
that qualitatively very similar results were obtained if one
chooses $r_n$ a bit smaller than this choice (larger values of
$r_n$ are physically hardly significant, since then the particles
are too weakly bound, due to the rapid fall-off of the LJ
potential). The same distance $r_{n}$ is considered also for the
analysis of the number of contacts. Therefore, a pair of monomers
being an absolute distance  less than $r_{n}=1.5$ apart define a
``contact''. Then, the numbers presented in this manuscript denote
the average number of neighbors per monomer.

In the case of properties as it is, for example, the number of
contacts, one just considers the thermal average. However, to
define properties related to the fluctuating number of clusters
one should also take into account averages of quantities $A$ that
depend on these fluctuations, i.e.,
\begin{equation}\label{eq:5}
\bar{A} = \sum \limits _{N_{cl}}P(N_{cl}) A(N_{cl}).
\end{equation}

\begin{figure}[h]\centering
\includegraphics[height=8cm, angle=270]{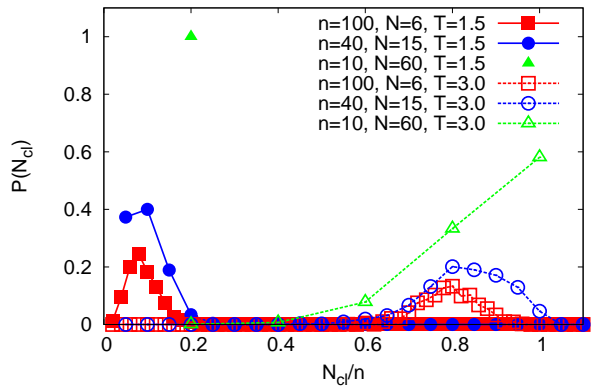}
\caption{Probability distribution $P(N_{cl})$ of the number of
clusters plotted versus $N_{cl}/n$ for combinations of $n$ and $N$
that correspond to the snapshots of Fig.~\ref{fig:2}. Results are
shown for two temperatures, $T=1.5$ and $T=3.0$. Lines are guides
for the eye.} \label{fig:3}
\end{figure}
As an example, Fig.~\ref{fig:3} shows data of the distribution in
the number of clusters $N_{cl}$ for several choices of $n$, $N$. The
extreme choices would be that every block forms a separate cluster
($P(N_{cl})=1$ for $N_{cl}/n=1$, where A and B are always
phase separated), which is seen for none of the cases
presented in Fig.~\ref{fig:3} (the case $n=10$ at temperature
$T=3$ shows that this case is possible with a probability $\approx 0.6$),
and that all blocks of A-type monomers form a cluster with the
blocks of B-type forming another cluster ($P(N_{cl})=1$
for $N_{cl}/n=2/n$). The latter case is taking place for $n=10$ at
$T=1.5$, where we see a single point showing that all blocks of
type A are always belonging to the same cluster and all blocks of
type B to another cluster. The other cases presented in
Fig.~\ref{fig:3} are intermediate ones, where we have a symmetric
variation in the number of clusters around an average value.

\section{Results and discussion}
\label{sec:3}

\begin{figure}[h]\centering
\includegraphics[height=8cm, angle=270]{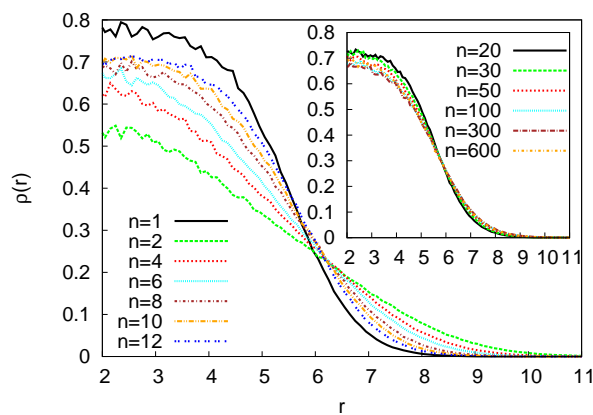}
\caption{Density profile $\rho(r)$ plotted versus the radial
distance $r$ from the center of mass of the chain at temperature
$T=1.5$. Note the pronounced differences occurring for small $n$, whereas the
differences for $n>12$ are minor.} \label{fig:4}
\end{figure}

The incompatibility between A and B blocks can increase the
overall chain dimensions of the chains compared to the homopolymer
chain of the same length under the same thermodynamic conditions.
A and B monomers prefer to be apart, increasing in this way the
overall size of the macromolecule. This can be readily checked if
we measure the overall density profile of the multiblock
copolymers relative to the center of mass for various cases of $n$
($n \propto 1/N$ for $nN=const$). In Fig.~\ref{fig:4} we present
this dependence on $n$ for $nN=600=const$ at a low temperature,
i.e., $T=1.5$. The total chain length ($nN=600$) is high enough to
discuss the different scenarios of microphase separation in these
macromolecules as the length of the blocks varies up to relatively
large values of $N$. Indeed, we can see from Fig.~\ref{fig:2} that
already for $N=60$ the formation of single clusters takes place.
Then we can focus on the variation with $n$ and $N$, with the
latter parameter being of course mostly important.

The first quantity we have computed is the overall monomer
density, mathematically expressed by the following formula
\begin{equation}\label{eq:6}
\rho(|\vec{r}|)= \langle  \sum \limits_{i=1}^{nN} \delta(
\vec{r}-\vec{r}_{c}-\vec{r}_{i}) \rangle,
\end{equation}
where $\delta(\vec{x})$ is the Dirac delta function, $\vec{r}_c$
the position of the center of mass of the whole chain, and
$\vec{r}_i$ the positions of all monomers, irrespective of their
type (A or B). The angle brackets denote an average over all
conformations as usual. Figure~\ref{fig:4} shows the results for
the overall monomer density, where $n$ is varied, but $nN$ is kept
constant and equal to $600$. These results refer to a temperature
$T=1.5$. We can clearly see that the homopolymer chain ($n=1$) of
length ($nN=600$) obtains always the most compact conformations,
if we compare it with the respective cases of multiblock chains.
The case of $n=2$ exhibits the greatest difference from the
homopolymer case, with the former showing bigger dimensions of the
chain, due to the formation of a large A-B interface. As $n$
increases up to $n=12$ the multiblock chains gradually exhibit
more compact structures, but they are always more swollen than the
homopolymer case. For the range of $n$ between $20$ and $600$
shown as an inset in Fig.~\ref{fig:4} we see that the differences
in the density profiles for different multiblock chains are rather
small. However, for this range of values ($n=20-600$) we observe a
monotonic behavior, where the chains obtain less compact
structures with increasing $n$. This is clearly due to the
increase of the unfavorable interactions between A and B monomers.
Actually, we note here that the increase of $n$ should result in
breaking of the spherical symmetry, but such effects will not be
discussed in this manuscript. Here, we are only concerned with the
phase separation of multiblock copolymers under poor solvent
conditions.
\begin{figure}[h]\centering
\includegraphics[height=8cm, angle=270]{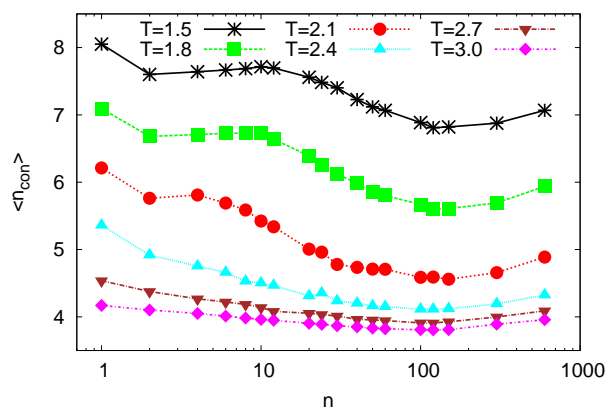}
\caption{Average number of contacts $\langle n_{con} \rangle$ per
monomer plotted versus the number of blocks $n$ for different
temperatures distinguished with different symbols and line
formats. Note the logarithmic scale in the $x$-axis and the
different regions that appear reflecting the different behavior of
multiblock copolymers with different $n$.} \label{fig:5}
\end{figure}
When the density profiles of these systems at significantly higher
temperatures (and closer to the Theta temperature, but still under
poor solvent conditions) are plotted, somewhat smaller differences
occur and the behavior observed at high $N$ is slightly different.
These effects will be explained below as we will discuss in detail
the effect of the temperature.

As $n$ increases, it is natural that the number of contacts
between A and B monomers increases and then the number of contacts
A-A and B-B will decrease. Therefore, we are interested in the
overall number of contacts per monomer irrespective of whether a
monomer is of type A or B. These results are presented in
Fig.~\ref{fig:5} for different temperatures. We observe that
$n_{con}$ is higher in the case of the homopolymer chain. This
confirms our understanding for the homopolymer case, already seen
in the calculation of the overall profile. The homopolymer chain
is always the most compact globule at this low temperature $T=1.5$
and of course no multiblock chain can obtain such a compact
structure that the overall contacts are mostly favored. Moreover,
the results of Fig.~\ref{fig:5} show that this is true for all
temperatures up to $T=3.0$. Therefore, the effect of
incompatibility between A and B is rather weakly present at higher
temperatures, as expected. These effects would become less
noticeable as the temperature increases and becomes considerably
higher than the Theta temperature they should be unnoticeable. For
temperature $T=1.5$ and small $n$ the curve has a plateau from
$n=2$ to $n=10$. This result is surprising, since it shows that
the globule at this low temperature has bigger dimensions, but the
overall number of contacts stays almost constant as the effect of
the solvent incompatibility sets in strongly.
\begin{figure}[h]\centering
\includegraphics[height=8cm, angle=270]{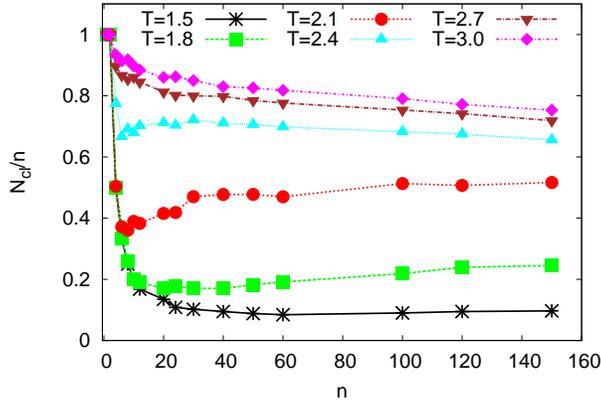}
\caption{Average number of clusters $N_{cl}$ divided by
the total nunber of blocks ($n$) versus $n$ for different
temperatures, as indicated.} \label{fig:6}
\end{figure}
Although there are obvious differences in the density profiles of
Fig.~\ref{fig:4}, the multiblock chain for small $n$ rearrange in
such a way that the number of contacts stays almost constant. For
values higher than $n=12$ the number of contacts decreases and in
the case of very high $n$ ($n>100$) we see a small increase. In
the latter case, we notice that the number of monomers per block
is small ($N=4$ to $N=1$). When the temperature increases, we
see that the plateau for small $n$ becomes shorter and for
temperatures $T>2.4$ has completely disappeared. This shows
that for $T>2.4$ the effect of the temperature on the behavior of
the multiblock chain is minor and its behavior is defined by the
incompatibility between species A and B. We conclude that although
the number of contacts A-B increases with increasing $n$, the
effect of the temperature (when $T<2.4$) keeps the number of
overall contacts almost constant for $N>30$. At temperature
$T>2.1$ the curves show a monotonous decrease in the number of
contacts and a slight increase for high $n$. At temperature
$T=3.0$ the variation of $n$ in multiblock chains has a very small
effect and the difference between $n=1$ and $n=2$ is rather
indistinguishable.

A better understanding of all the above effects can be achieved in
terms of the analysis of the formed clusters from blocks of the
same type. In Fig.~\ref{fig:6} we have computed the average number
of clusters with blocks of type A together with the different
clusters of type B. Of course, when $n=1$ we have only one cluster
with one type of monomers, and for $n=2$ we have always one
cluster with monomers of type A and one cluster with monomers of
type B and each cluster comprises only one block. For small $n$
the ratio $N_{cl}/n$ should be one. At temperature $T=1.5$,
$N_{cl}/n$ stays zero up to $n=12$. This plateau regime starts at
smaller $n$ as the temperature increases, in accordance with the
results of our previous figures. For $n>30$ the ratio $N_{cl}/n$
stays almost constant showing that the effect of the temperature
takes over the effect of the incompatibility between the monomers
of different type.
\begin{figure}[h]\centering
\includegraphics[height=8cm, angle=270]{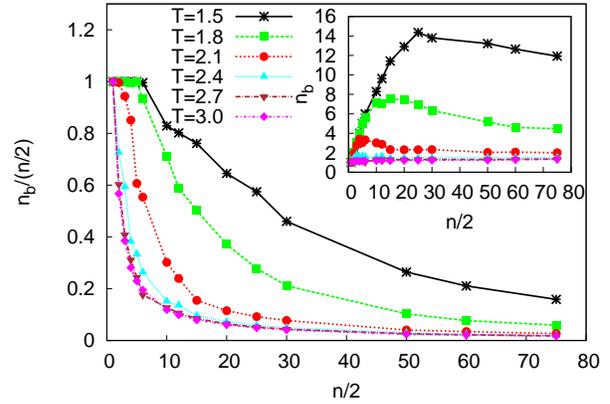}
\caption{Number of blocks $n_{b}$ in each cluster (A or B) divided with
the number $n/2$ of A (or B) blocks versus this number ($n/2$).
The inset shows the number of blocks in a cluster (A or B) versus
$n/2$.} \label{fig:7}
\end{figure}
At higher temperatures, i.e., $T=1.8$ and $T=2.1$, $N_{cl}/n$
slightly increases. For even higher temperatures a monotonic
behavior is observed. The highest variations are observed for high
$N$, where the ratio $N_{cl}/n$ decreases rather sharply.

We can go one step further in order to understand the above
phenomena. We analyze the number of blocks $n_{b}$ of type A (or B) in a
cluster of type A (or B), respectively. Then, the number of blocks
A is $n/2$. Correspondingly the number of blocks with monomers of
type B is $n/2$. Due to the symmetry of our model we expect the
same results for the analysis of clusters with blocks of type A
and of clusters with blocks of type B. Indeed, our results show
that this is true confirming the equilibration procedure, which
was adopted in this study. In Fig.~\ref{fig:7} we show such plots
where the number of blocks of a cluster is plotted versus the
total number of A (or B) blocks of the multiblock chain. It is now
clearly seen the range of $N$ where all blocks of the same type
belong to the same single cluster. This happens when
$n_{b}/(n/2)=1$. When the temperature is rather low ($T=1.5$) this
plateau extends up to values $n=20$, which corresponds to block
length of $N=30$. This means that the multiblock chain is
separated in two clusters, that is one of type A and another of
type B. This case then corresponds to the chain of $n=20$, and this
explains the results of Figs.~\ref{fig:5} and \ref{fig:6}. When
the temperature is lower, the number of blocks in the clusters is
higher and therefore the curve corresponding to $T=1.5$ shows a
smooth gradual decrease. More abrupt is the change in higher
temperatures where the ratio $n_{b}/(n/2)$ decreases abruptly.
\begin{figure}[h]\centering
\includegraphics[height=8cm, angle=270]{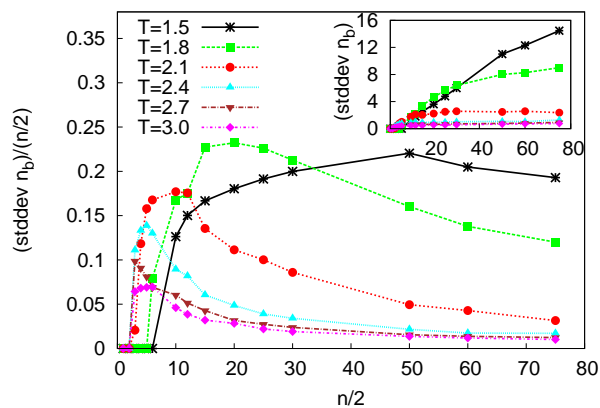}
\caption{Standard deviation in the number of blocks in each
cluster (A or B) divided with the number $n/2$ of A (or B) blocks
that the multiblock chain contains versus this number ($n/2$). The
inset shows the standard deviation in the number of blocks in a
cluster (A or B) versus $n/2$.} \label{fig:8}
\end{figure}
Then the plateau that follows with increasing $n$ shows the range
of $n$ that the enthalpic interactions take over. For $T>2.4$ the
chains exhibit almost the same behavior with a sharp decrease at
small $n$ and a plateau at higher $n$. This shows that at high
temperatures and high $n$ hardly ever two blocks of the same type
come together to form a cluster. Then the ratio $n_{b}/(n/2)
\approx 1/(n/2)$. However, at temperature $T=1.5$ this ratio has
the value $n_{b}/(n/2)\approx0.2$ showing that the formation of
clusters with more than two blocks is possible. The inset shows
the above results in a different representation. Then the linear
curve ($y=ax$) denotes the regime that the ratio $n_{b}/(n/2)=1$.
It is also seen that this regime becomes smaller as the
temperature is increased. Then, at low temperatures, the
incompatibility between the monomers of different type becomes
significant. This is not seen at high temperatures
close to Theta, where the effect of the solvent quality and the
enthalpic interactions rather cancel each other.

The number of blocks contained in a cluster is not the same, but
it is a quantity that fluctuates, as we have already seen from
Fig.~\ref{fig:3}. Then Fig.~\ref{fig:8} shows these fluctuations
of the number of blocks in a cluster. For small values of $n$ (or
high $N$) such fluctuations should not exist, since there are only
two clusters, each one with blocks of monomers of different type.
For intermediate values of $n$ the fluctuations in the number of
blocks in the clusters become rather high. The blocks of type A
try to rearrange in such way that they form a single cluster, but
the number $n$ is high enough to hinder such effect. For $n$ in
the regime of strong fluctuations the formation of single clusters
containing all blocks of the same type is possible. For higher $n$
the curves show a monotonic smooth behavior and the enthalpic
interactions take over. It is also noticeable that the most
pronounced fluctuations are observed in lower temperatures, where
the effect of the incompatibility between different monomer types
and the effect of the solvent incompatibility compete.
\begin{figure}[h]\centering
\includegraphics[height=8cm, angle=270]{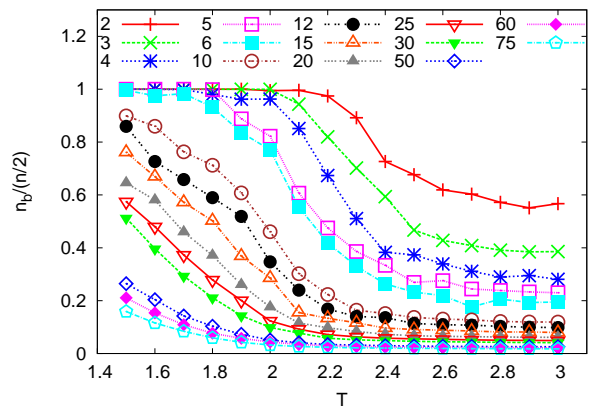}
\caption{Number of blocks $n_{b}$ in each cluster (A or B) divided with
the number $n/2$ of A (or B) blocks that the multiblock chain
contains versus temperature.} \label{fig:9}
\end{figure}
At $T=1.5$ the fluctuations persist over high values of $n$. This
is clearly seen from the inset of Fig.~\ref{fig:8}. Also, as $N$
becomes smaller, the fluctuation in the number of blocks in a
cluster is higher.

Interesting is the dependence on the temperature for this
particular case. From Fig.~\ref{fig:9} we can clearly distinguish
three different cases. For small values of $n$ ($2<(n/2)<6$) the
formation of a single cluster is possible up to temperatures
$T=1.8$. In this case the blocks of the same type belong always to
the same cluster. This behavior is weakened as $n$ increases, or
correspondingly $N$ decreases. The second case is that for
$10<(n/2)<30$, where the formation of clusters containing all
blocks of the same type is possible. Of course, this is more
probable when $N$ is higher and the temperature is low. As the
temperature is increased, we obtain a strong decrease in the
number of blocks per cluster and at high temperatures $T>2.3$ we
obtain a plateau where $n_{b}/(n/2)\approx 2/n$. When $n$ is high
the dependence on the temperature is rather small and the behavior
of the chain is defined by the incompatibility between A and B
monomers. Of course for the latter case the formation of single
clusters with all blocks of the same type is impossible.

\begin{figure}[h]\centering
\includegraphics[height=8cm, angle=270]{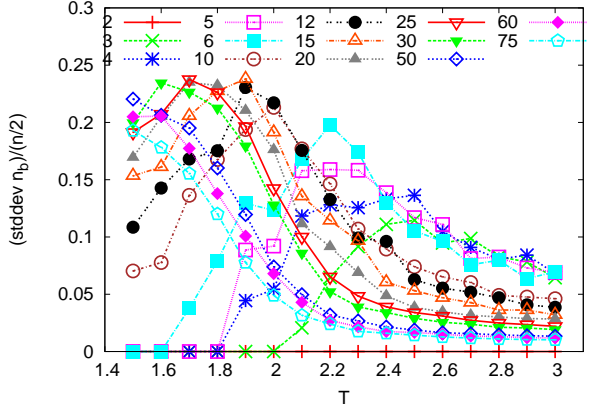}
\caption{Standard deviation in the number of blocks $n_{b}$ in each
cluster (A or B) divided with the number $n/2$ of A (or B) blocks
that the multiblock chain contains versus temperature.}
\label{fig:10}
\end{figure}

Figure~\ref{fig:10} shows the dependence of the fluctuation in the
number of blocks per cluster. The stronger fluctuations are
observed at lower temperatures (depending on $n$) in the range
between $T=1.7$ and $T=2.4$. As $n$ increases this peak of the
higher fluctuations is shifted to the left at lower temperatures.
The results of Figs.~\ref{fig:9} and \ref{fig:10} show that at low
temperatures the formation of single clusters is possible, but the
same time the fluctuations in the number of blocks in the cluster
are rather high.

\begin{figure}[h]\centering
\includegraphics[height=8cm, angle=270]{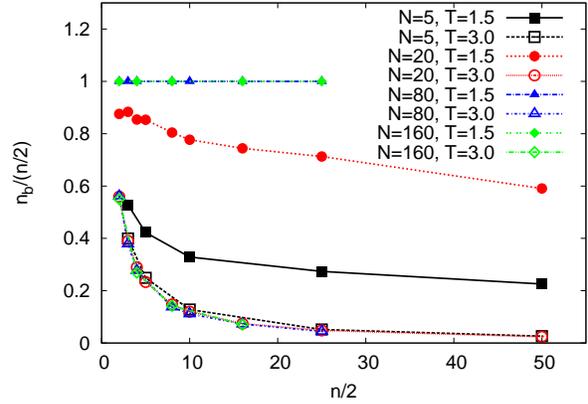}
\caption{Number of blocks $n_{b}$ per cluster divided with the number of
blocks of type A (or B) versus this number of blocks $n/2$.
Results concern different block lengths $N$ and temperatures $T$.}
\label{fig:11}
\end{figure}

For block copolymers we know that the incompatibility between
different monomers is expressed by the term $\chi N$. Here, we
discuss the effect of the block length $N$. We choose three
different chain lengths, which correspond to low $N$,
intermediate, and high enough, where always the blocks of the same
type form a single cluster. Then we look for the dependence on
$n$. The results of Fig.~\ref{fig:11} confirm the above argument
that the phase separation depends strongly on $N$, as expected,
and rather weakly on $n$ for a collapsed globule. Of course the
influence of the surface energy ($ \propto R^2$, with $R$ being
the radius of the almost spherical collapsed object) is a
combination of the above two parameters. $n$ has such effect with
$N$ having the additional effect of controlling the
incompatibility between A and B monomers. When $N$ is high ($N=80$
or $160$), the blocks are joined together into a single cluster
with monomers of the same type at temperature $T=1.5$. Therefore,
for this case no dependence on $n$ is observed. However, when $n$
becomes larger than some value, the formation of more than one
cluster takes place. Then the overall length of the multiblock
chain is very long and difficult to study with simulation techniques. Thus,
such effect is not discussed in this manuscript and is not
systematically studied.
For intermediate $N$ ($N=20$) and small $n$ the
formation of a single cluster is still possible, but this length
is not high enough to ensure strong incompatibility between the
blocks with monomers of different type. Somewhat richer behavior
is expected as it has already been discussed, with strong
fluctuations in the number of blocks per cluster. For $N=5$ the
formation of a single cluster is not seen and the number of blocks
per cluster and per block decreases reaching a plateau value at
high $n$. For $T=3.0$ we see that that the different block length
does not play any role. The three different block lengths have
collapsed into a single curve ($\chi N \propto \Delta
\epsilon/T$). This is the indication that the behavior of our
multiblock chain depends strongly on the effect of the
temperature. At high temperature also the curve reaches a plateau,
that is for $(n/2)>25$.

\begin{figure}[h]\centering
\includegraphics[height=8cm, angle=270]{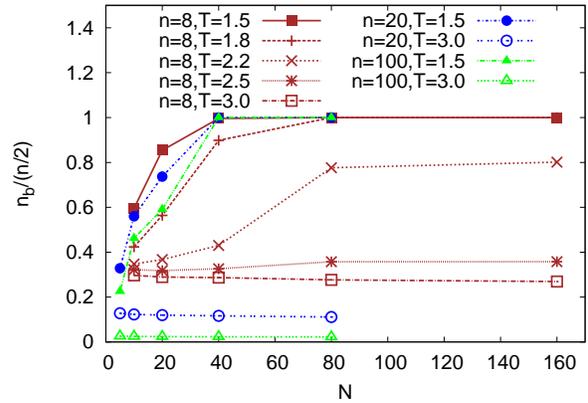}
\caption{Same as Fig.~\ref{fig:11}, but the dependence on
the block length $N$ is shown. The curves correspond to different
numbers of blocks $n$ and temperatures $T$.} \label{fig:12}
\end{figure}

Figure~\ref{fig:12} shows the dependence of the number of blocks
per cluster on $N$. For $(n/2)=4$ and at low temperature we have
the formation of a single cluster for $N>20$. As the temperature
increases this happens for higher values of $N$. At $T=2.2$ the
curve takes gradually a constant value. For temperature $T=3.0$,
which corresponds roughly to the Theta temperature, there is no
dependence on $N$, as it is expected. This is also observed for
$n=20$ and $n=100$. Comparing the results for different $n$ at the
low-temperature regime, we see that the phase separation depends
rather strongly on $N$ and rather weakly on $n$. We should note
here that all phase changes occur gradually due to the presence of
thermal fluctuations in a collapsed object with a high ratio of
surface to bulk energy.

\section{Conclusions}
\label{sec:4}

In this manuscript we have presented a detailed study of the phase
behavior of multiblock copolymers varying the number of blocks
$n$, the length of the blocks $N$, and the solvent quality by
variation of the temperature. We have described the different
types of phase separation appearing in linear multiblock
copolymers and we have discussed in detail the effect of these
parameters. Although the temperature (solvent incompatibility) and
the block length $N$, as in the bulk, control the phase separation
between blocks of different type (for given $\Delta \epsilon$),
the effect of the number of blocks $n$ is also important due to
the increase of the surface free energy.
In the range of values that
have been discussed here, very reach behavior is observed for
intermediate values of $n$, $N$ and $T$, where the fluctuations in
the number of blocks per cluster are pronounced. In this case the
interplay between enthalpic and entropic effects compete resulting
in different types of microphase separation in the collapsed
states. The behavior of multiblock copolymers can be parallelized
with that of various biological macromolecules which are formed by
periodically repeated structural units (``monomers'') along their
chain. Our study provides a first step towards the understanding
of the phase separation of such complex systems.

\textbf{Acknowledgments} One of us (P.E.T.) gratefully
acknowledges financial support through a Max Planck fellowship
awarded by the Max Planck Institute for Polymer Research.

\balance
\footnotesize{

}

\end{document}